\documentclass[12pt]{scrartcl}

\usepackage{moreverb}
\usepackage[colorlinks,bookmarksopen,bookmarksnumbered,citecolor=red,urlcolor=red]{hyperref}
\usepackage{natbib}
\usepackage{animate}
\usepackage{multimedia}
\usepackage{amsmath}
\usepackage{amsthm}
\usepackage{amssymb}
\usepackage{bm}
\usepackage{color}
\usepackage{graphicx}
\usepackage{enumerate}
\usepackage{microtype}
\usepackage{nicefrac}
\usepackage{booktabs}
\usepackage[T1]{fontenc}
\usepackage[font=small,labelfont=bf,tableposition=top]{caption}
\usepackage[toc,page]{appendix}
\usepackage{algorithm}
\usepackage{algorithmic}
\usepackage[affil-it]{authblk}
\usepackage[flushmargin,ragged]{footmisc} 

\DeclareCaptionLabelFormat{andtable}{#1~#2  \&  \tablename~\thetable}

\definecolor{darkred}{rgb}{0.5,0.1,0.1}
\hypersetup{
	colorlinks   = true,   
	urlcolor     = darkred,    
	linkcolor    = darkred,    
	citecolor    = darkred    
}

\DeclareMathOperator{\diag}{diag}
\newcommand{\bs}[1]{\boldsymbol{#1}}
\providecommand{\keywords}[1]{\textbf{\textit{Keywords: }} #1}

\title{A determinant-free method to simulate the parameters of large Gaussian fields}
\date{27 June 2017}
\author[1]{Louis Ellam\thanks{Email: \href{mailto:l.ellam16@imperial.ac.uk}{l.ellam16@imperial.ac.uk}}}
\author[2]{Heiko Strathmann}
\author[1,3]{Mark Girolami}
\author[4]{Iain Murray}
\affil[1]{Department of Mathematics, Imperial College London, London, SW7 2AZ, UK}
\affil[2]{Gatsby Unit for Computational Neuroscience, University College London, London, W1T 4JG, UK}
\affil[3]{The Alan Turing Institute, The British Library, London, NW1 2DB}
\affil[4]{School of Informatics, University of Edinburgh, Edinburgh, EH8 9AB}

\begin{document}

\maketitle

\begin{abstract}
We propose a determinant-free approach for simulation-based Bayesian inference in high-dimensional Gaussian models.
We introduce auxiliary variables with covariance equal to the inverse covariance of the model. The joint probability of the auxiliary model can be computed without evaluating determinants, which are often hard to compute in high dimensions.
We develop a Markov chain Monte Carlo sampling scheme for the auxiliary model that requires no more than the application of inverse-matrix-square-roots and the solution of linear systems. These operations can be performed at large scales with rational approximations.  We provide an empirical study on both synthetic and real-world data for sparse Gaussian processes and for large-scale Gaussian Markov random fields.
\end{abstract}


\section{Introduction}
Linear Gaussian models are one of the most fundamental tools for statistical analysis, with diverse applications such as spatial modelling, uncertainty quantification, dimensionality reduction, and time-series modelling, e.g.\ speech recognition or even analysing high-dimensional recordings of neural activity \citep{roweis1999unifying, rosti2004linear, macke2011empirical}.
A set of $n$ observations, $\bs{y}$, is assumed to  be generated from  a set of latent Gaussian variables, $\bs{x}$, which are subject to a linear transformation $\bs{A}$  and corrupted by independent Gaussian noise $\bs{\epsilon}$.
Popular examples include non-parametric regression with Gaussian processes \citep[GPs,][]{Rasmussen06gaussianprocesses} or modelling spatial phenomena using Gaussian Markov random fields \citep[GMRFs,][]{rue2005gaussian, lindgren2011explicit}.
GPs are often based on full covariance matrices, which means none of the variables are marginally independent. In contrast GMRFs directly encode conditional independence relationships via zero elements in the precision matrix, which as a result is often sparse.

We focus on Bayesian posterior inference for linear Gaussian models. It is possible to obtain closed form expressions for key quantities such as the marginal likelihood and posterior predictions.  However, computing these quantities in practice is difficult for large problems. For $n$ modelled variables, an evaluation of the (normalised or unnormalised) probability density function costs $\mathcal{O}(n^3)$ time and $\mathcal{O}(n^2)$ memory.
These costs arise from the solution of a linear system and the evaluation of a matrix determinant.
There is a large literature on computationally motived approximations to the model itself, for example via sub-sampling or low-rank approximations \citep{quinonero2005unifying}, however, exact Bayesian inference for large models remains a challenging problem.

We can compute some properties of models with sparse covariance or precision matrices
using two families of numerical linear algebra methods.
These methods exploit the fact that large sparse matrices can be stored, and their products with vectors can be computed.
Krylov methods, such as the linear solver conjugate gradient and multi-shift variants \citep{simpson2008krylov, jegerlehner1996krylov}, can solve linear systems with just matrix-vector products.
Rational approximations reduce the computation of matrix functions, such as the determinant, inverse, or square root, to solving a family of linear equations -- with error guarantees that are straight-forward to control and can even be set to floating point precision \citep{higham2013matrix, kennedy2004approximation}.
A combination of these methods has been successfully applied to sample from high-dimensional Gaussian distributions \citep{aune2013iterative}, and to perform maximum likelihood inference by estimating the log-determinant term \citep{aune2014parameter}.  { A number of numerical approximations are available for estimating the log-determinant term, for example, those discussed by \citet{bai1996some}, \citet{han2015large}, and \citet{saibaba2016randomized}.}  Recently, an inversion-free approach for inferring point-estimates of covariance parameters of large-scale Gaussian distributions was proposed by \citet{anitescu2016inversion}.
The approach is based on an alternative objective function than maximum likelihood, while asymptotically converging to the same solution.

Bayesian inference of the covariance parameters, however, remains largely an open problem.
Markov chain Monte Carlo (MCMC) methods are `asymptotically exact', but most previous work has
limited scalability as the methods need to compute determinants \citep[e.g.,][]{murray2010}.
\citet{lyne2015russian} realised that the log-determinant estimator by \citet{aune2014parameter} is unbiased, and can therefore be combined with the `exact-approximate' framework within a pseudo-marginal MCMC algorithm \citep{Andrieu2009a} -- as also used by \citet{filippone2014pseudo}.
A complication is the need to transform an unbiased estimator of the log-determinant into an unbiased estimator of the determinant itself, which \citet{lyne2015russian} achieved via random truncation (Russian roulette) of the infinite series expression of the exponential function.
Apart from theoretical issues around the existence of positive unbiased estimators for Russian roulette \citep{jacob2015nonnegative}, unfortunately, for a real-world model of global ozone distributions \citep{lindgren2011explicit}, the combination of Russian roulette and pseudo-marginal MCMC turned out to be too fragile and the resulting Markov chain in practice did not converge reliably.

In this paper we introduce an alternative MCMC approach to perform asymptotically exact inference on large-scale Gaussian models. We introduce auxiliary variables so that the joint distribution of the model and auxiliary variables contains no determinants. In return for removing determinants we need to update the auxiliary variables, which can be performed with the application of inverse-matrix-square-roots.  In the case that the inverse-matrix-square-root of the covariance is unknown, we use rational approximations in the spirit of \citet{aune2013iterative} to perform tuning-free updates.  Our scheme is considerably simpler than the approach taken by \citet{lyne2015russian}, and by avoiding the pseudo-marginal framework there is no need to tune the internal unbiased estimators.  Our approach scales well to high-dimensional models, provided that the application of an inverse-matrix-square-root can be carried out reliably.  In the case of a poorly-conditioned model, in which the inverse-matrix-square-root is unknown, the underlying Krylov method may converge slowly, or not at all to within the desired tolerance.

In the remainder of the paper, we give details on the linear Gaussian model itself and the computational challenges, here for the case of using MCMC for posterior inference.
We introduce our auxiliary model, which avoids the need to compute matrix determinants, and describe a sampling scheme for the resulting joint distribution -- including some necessary background on rational approximations.
Finally, we empirically study the method on some toy and real-world examples.
For models with well-behaved covariance or precision matrices, our determinant-free method can outperform MCMC using standard Cholesky factorisations.

\section{Background}
We consider linear models of the form
\begin{equation}
\label{eq:linear_gaussian}
\bs{y} = \bs{A x} + \bs{\epsilon}.
\end{equation}
Here, observations $\bs{y} \in \mathbb{R}^n$ are a linear transformation $\bs{A} \in \mathbb{R}^{n \times m}$ of independent Gaussian latent variables,
\begin{equation}
\label{eq:latent_variables}
\bs{x} \in \mathbb{R}^m\sim \mathcal{N}(\bs{\mu_\theta}, \bs{\Sigma_\theta}),
\end{equation}
with independent Gaussian additive noise\footnote{For simplicity we include $\tau$ in the parameters $\bs{\theta}$, although $\tau$ doesn't effect the covariance of the latents $\bs{x}$.}
$\bs{\epsilon} \in \mathbb{R}^{n}$ and covariance matrix $\tau^{-1}\bs{I}$.  Since the model is linear and Gaussian, the marginal likelihood, $p(\bs{y} | \bs{\theta} )$, is also Gaussian, i.e.\ 
\begin{align}
\label{eq:marginal_likelihood}
\bs{y} | \bs{\theta} &\sim 
\mathcal{N}(\bs{A \mu_\theta}, \bs{S_\theta})
\end{align}
with covariance $\bs{S_\theta}=\tau^{-1} \bs{I} + \bs{A}\bs{\Sigma_\theta}\bs{A}^\top$.

Two common examples with this set-up are finite-dimensional realizations of a Gaussian process (GP) and Gaussian Markov random field (GMRF) models, often with dense covariance and sparse precision matrices, respectively.

For these models to be of any practical use, it is necessary to determine suitable parameter values $\bs{\theta}$.
In a Bayesian setting, we define a prior density $p(\bs{\theta})$  and combine it with the likelihood in~\eqref{eq:marginal_likelihood}. Bayes' theorem induces the posterior
\begin{equation}
\label{eq:mot_posterior}
p( \bs{\theta} | \bs{y}) \propto p(\bs{y} | \bs{\theta} ) p (\bs{\theta}),
\end{equation}
which is intractable for most non-trivial applications.  While there are many approximate inference schemes to explore the posterior~\eqref{eq:mot_posterior}, e.g.\ variational methods, we here focus on simulation via Markov chain Monte Carlo (MCMC), which has the advantage of being asymptotically exact.
MCMC methods seek to generate samples $\{\bs{\theta}^{(i)} \}_{i=1,2,\dots} \sim p(\bs{\theta} | \bs{y})$ to represent the posterior.
These samples can be used to estimate the posterior expectation of arbitrary functions, such as the posterior mean.
In this paper we will use the standard Metropolis--Hastings (MH) algorithm.

\subsection{Determinants in high dimensions}
To evaluate the likelihood in~\eqref{eq:marginal_likelihood}, the standard approach is to compute the Cholesky factorisation $\bs{S_\theta} = \bs{R_\theta}^\top\bs{R_\theta}$, where $\bs{R_\theta}$ is upper triangular.
This factorisation allows any linear system of the form $\bs{S_\theta}^{-1}\bs{x}$ to be solved via cheap back-substitution, and provides the log-determinant as
\begin{equation*}
\log |\bs{S_\theta}| = 2 \sum_{i=1}^n \log( \diag({\bs{R_\theta}})_i).
\end{equation*}
For sparse high-dimensional matrices $\bs{S_\theta}$, however, this approach is unsuitable, as even storing the Cholesky factorisation is infeasible. Cholesky factors suffer from a so called \emph{fill-in} effect and are generally not sparse.  Currently standard computers struggle to hold Cholesky factors in memory from around $n\!=\!\text{10,000}$.
Examples of large sparse covariance matrices are those of Gaussian processes with compactly supported covariance functions. The Cholesky factor of sparse precision matrices from GMRF models can also be used.

Unbiased estimates of log-determinants can be obtained with rational
approximations and Krylov subspace methods \citep{aune2014parameter}, although
at greater expense than solving linear systems. Using these approximations to
construct an MCMC method is complicated \citep{lyne2015russian}.

\section{Methodology}

We augment the state space of the linear Gaussian model by introducing auxiliary variables
\begin{equation}
\label{eq:auxiliary_variables}
\bs{z} | \bs{\theta} \sim \mathcal{N}(\bs{0}, \bs{S_{\bs{\theta}}}^{-1}).
\end{equation}
The resulting joint posterior is then given by the product of the posterior $p(\bs{\theta}|\bs{y})$ in~\eqref{eq:mot_posterior} and the distribution over the new auxiliary variables,
\begin{equation}
p(\bs{\theta}, \bs{z} | \bs{y}) 
\propto p(\bs{\theta}) e^{ -\frac{1}{2} (\bs{y}-\bs{A}\bs{\mu_\theta})^\top \bs{S}_{\bs{\theta}}^{-1} (\bs{y}-\bs{A}\bs{\mu_\theta})
	- \frac{1}{2} \bs{z}^\top \bs{S_{\bs{\theta}}} \bs{z}}.
\label{eq:aug_joint}
\end{equation}
The normalising terms, i.e.\ determinants, of the original model and the augmented variables cancel, 
\begin{align*}
\frac{1}{\int\int p(\bs{\theta} | \bs{y} )p(\bs{z}|\bs{\theta})d\bs{z}d\bs{\theta}}\propto \frac{1}{|\bs{S_\theta}|^{\frac{1}{2}} |\bs{S_\theta}|^{-\frac{1}{2}}}  \propto 1,
\end{align*}
and~\eqref{eq:aug_joint} is left with only the quadratic forms for the latent and auxiliary Gaussian variables.  The original posterior in~\eqref{eq:mot_posterior} is restored by marginalizing over $\bs{z}$,
\begin{equation*}
p(\bs{\theta}| \bs{y}) = \int p(\bs{\theta}, \bs{z} | \bs{y}) \, d\bs{z}.
\end{equation*}
In other words: in order to obtain posterior samples, we can sample from the augmented distribution and subsequently discard the auxiliary variables.  Our MCMC scheme alternates between two updates:
\begin{enumerate}
	\item Update the auxiliary variables $\bs{z}$ keeping $\bs{\theta}$ fixed using an MCMC method for target density~\eqref{eq:auxiliary_variables}.
	\item Update the parameters $\bs{\theta}$ keeping $\bs{z}$ fixed using an MCMC method with target proportional to~\eqref{eq:aug_joint}.
\end{enumerate}
We next specify how we implement these updates.

\subsection{Model parameter updates}
The model parameters can be updated with standard Metropolis--Hastings (MH) updates.
We use a preliminary run to estimate the covariance of the posterior $p(\bs{\theta}|\bs{y})$.  In our experiments we use a random walk proposal that is tuned to achieve an acceptance rate in the range of $20$--$40\%$ \citep{rosenthal2011optimal}.

\subsection{Auxiliary variable updates}
\label{sec:methods_update_aux}

The auxiliary variables $\bs{z}$ could potentially be updated with MH or Hamiltonian Monte Carlo methods. However, the tuning of these updates would be challenging.  Instead we perform Gibbs updates, directly resampling the auxiliary $\bs{z}$ from its conditional distribution $\mathcal{N}(\bs{0}, \bs{S_{\bs{\theta}}}^{-1})$.  In the case that the inverse-matrix-square-root of covariance is unknown, we can't use a Cholesky factorization to perform this update for large scale problems.  Instead we employ a technique from numerical linear algebra, a combination of rational approximations and Krylov sub-space methods \citep{aune2014parameter, kennedy2004approximation, higham2013matrix, jegerlehner1996krylov}.
These methods, outlined in the next section, are able to compute the product of the square-root-inverse of a matrix with arbitrary vectors -- only requiring matrix-vector products involving the matrix itself.
In particular, we can sample from high-dimensional Gaussian distributions as long as we can quickly apply the covariance or precision matrix to a vector, and that the matrix is reasonably well-conditioned.  We now distinguish the cases where we have direct access either to the sparse latent covariance matrix $\bs{\Sigma_\theta}$ from equation~\eqref{eq:latent_variables}, or to the precision $\bs{Q_\theta} = \bs{\Sigma_\theta}^{-1}$.

\paragraph{Sparse covariance}
For the case where we have direct access to a sparse latent covariance matrix $\bs{\Sigma_\theta}$, we directly compute $\bs{S_\theta}^{-\frac{1}{2}} \bs{w}$ for a standard normal $\bs{w}$, i.e.\  we generate a sample with desired auxiliary covariance $\bs{S_\theta}^{-1}$, via only using matrix-vector products of the form $\bs{S_\theta}\bs{w}=\tau^{-1} \bs{w} + \bs{A}\bs{\Sigma_\theta}\bs{A}^\top\bs{w}$, i.e.
\begin{align}
\label{eq:sample_aux_sparse_covariance}
\bs{w}\sim\mathcal{N}(\bs{0}, \bs{I}), \qquad \bs{z}\leftarrow \bs{S_\theta}^{-1/2}\bs{w}.
\end{align}

\paragraph{Sparse precision}
For the case where we know a sparse precision $\bs{Q_\theta}$, we create a set of `fantasy observations', $\tilde{\bs{y}}$ from the model.
For that, we first sample the latent variables $\tilde{\bs{x}}$ from~\eqref{eq:latent_variables} with covariance $\bs{Q_\theta}^{-1}$ by computing \smash{$\bs{Q_\theta}^{-\frac{1}{2}} \bs{w}'$}, using matrix-vector products of the form $\bs{Q_\theta}\bs{w}'$, where $\bs{w}'$ is standard Gaussian.
We then generate $\tilde{\bs{y}}=\bs{A}\tilde{\bs{x}}+\tau^{-1}\bs{w}'$.
Finally, we pre-multiply $\tilde{\bs{y}}$, which has covariance $\bs{S_\theta}$, with $\bs{S_\theta}^{-1}$, to get
\begin{align*}
\bs{S_\theta}^{-1}\tilde{\bs{y}}=\bs{S_\theta}^{-1} \bs{S_\theta}^\frac{1}{2}\bs{w}'=\bs{S_\theta}^{-\frac{1}{2}}\bs{w}',
\end{align*}
which has the desired auxiliary variable covariance $\bs{S_\theta}^{-1}$.
In practice, we employ the matrix inversion lemma:
\begin{align*}
\bs{S_\theta}^{-1}\tilde{\bs{y}}=\tau \tilde{\bs{y}} - \tau^2 \bs{A} ( \bs{Q_\theta} + \tau \bs{A}^\top\bs{A})^{-1} \bs{A}^\top \tilde{\bs{y}},
\end{align*}
which only requires a single additional linear solve in terms of matrix-vector products involving the known $\bs{Q_\theta}$.
In summary, given a sparse precision, we compute
\begin{align}
\label{eq:sample_aux_sparse_precision}
\bs{w}'\sim\mathcal{N}(\bs{0}, \bs{I}),  \qquad\bs{\epsilon}\sim\mathcal{N}(\bs{0}, \tau^{-1}\bs{I}),\nonumber\\
\tilde{\bs{x}}\leftarrow\bs{Q_\theta}^{-\frac{1}{2}}  \bs{w}', \qquad \tilde{\bs{y}}\leftarrow\bs{A}\tilde{\bs{x}}+ \bs{\epsilon}, \nonumber\\
\tilde{\bs{z}}\leftarrow \tau \tilde{\bs{y}} - \tau^2 \bs{A} ( \bs{Q_\theta} + \tau \bs{A}^\top\bs{A})^{-1} \bs{A}^\top \tilde{\bs{y}}.
\end{align}

\subsection{Rational approximations and Krylov methods}
We now briefly review the methodology required to solve the large sparse linear systems in~\eqref{eq:sample_aux_sparse_precision} and matrix-inverse-square-roots in~\eqref{eq:sample_aux_sparse_precision}.
Crucially, this is done without ever storing dense matrices, and only requires computing matrix-vector products $\bs{S_\theta}\bs{w}$ and $\bs{\Sigma_\theta}^{-1}\bs{w}$ respectively (we assume either the ability to compute $\bs{S_\theta}\bs{w}$ or that a sparse $\bs{\Sigma_\theta^{-1}}$ is given).
We mainly follow the approach taken by \citet{aune2013iterative}.
Rational approximations are used to construct matrix-vector products with matrix-inverse-square-roots, e.g.\ \smash{$\bs{S_\theta}^{-\frac{1}{2}}\bs{w}$}, by solving a family of sparse linear systems.
Krylov space methods may be used to solve linear systems using only matrix-vector products.

The key identity, the Cauchy integral formula \citep{kennedy2004approximation, clark2006rational}, relates a square matrix $\bs{A}$ and a function $f$ that is analytic on and inside the closed contour $\Gamma$ enclosing the Eigenvalues of $\bs{A}$ as
\begin{equation*}
f(\bs{A}) = \frac{1}{2 \pi i} \oint_{\Gamma} f(z)(z\bs{I} - \bs{A})^{-1} \, dz.
\end{equation*}
Right multiplying with vector $\bs{w}$ and applying a quadrature of the integral yields
\begin{equation}
\label{eq:rational_approximation}
f(\bs{A})\bs{w} \approx \sum_{i=1}^N \alpha_i (\bs{A} + \sigma_i \bs{I})^{-1} \bs{w},
\end{equation}
where we have to pick integration weights $\{\alpha_i\}_{i=1}^N$ and shifts $\{\sigma_i\}_{i=1}^N$ in order to ensure convergence.  For positive definite $\bs{A}$ and $f(\bs{A})=\bs{A}^{-\frac{1}{2}}$ and standard normal $\bs{w}$, Equation~\eqref{eq:rational_approximation} can be used to sample from a multivariate Gaussian with covariance $\bs{A}^{-1}$.

Integration weights and shifts in~\eqref{eq:rational_approximation} can be efficiently computed from the contour $\Gamma$, i.e.\ the smallest and largest Eigenvalue of $\bs{A}$, $m$ and $M$ respectively, in a standard way, e.g.\ \citet {aune2014parameter}.
To our knowledge, this is the best known rational approximation \citep{hale2008computing}.
Crucially, the Frobenius norm or the error in estimating $f(\bs{A})$ decays rapidly as
\begin{equation*}
\mathcal{O}\left(\exp\left({\frac{-2\pi^2N}{\ln(M/m)+3}}\right)\right),
\end{equation*}
and we can choose the number of quadrature points $N$ to reach a desired accuracy.
In practice, the contour $\Gamma$ should be chosen to enclose the smallest region that contains the spectral range of $\boldsymbol{A}$ for optimal efficiency.
We will see in the experiments, however, that we only need $N\approx 20$ quadrature points to reach floating point precision, e.g.\ $10^{-15}$ when using $m \!=\! 10^{-6}$ and $M \!=\! 10^6$.
Due to space constraints, we refer to the literature for further details.

\paragraph{Sparse (shifted family) linear systems}
The conjugate gradient algorithm can solve the $N$ linear systems required in~\eqref{eq:rational_approximation}, only requiring sparse matrix-vector products.
Conjugate gradient is guaranteed to converge after $n$ iterations for an $n$-dimensional system, where each matrix multiplication costs $\mathcal{O}(n^2)$.
Depending on the condition number of the underlying matrix, however, convergence up to a reasonable tolerance can happen at a fraction of $n$, and a preconditioning method can improve convergence rates \citep{benzi2002preconditioning, chow2014preconditioned}.
The $N$ linear equation systems in~\eqref{eq:rational_approximation} exhibit a special structure: only the diagonal term differs, arising from the various shifts $\{\sigma_i\}_{i=1}^N$.
Shifted family Krylov methods can solve these systems simultaneously at the cost of a \emph{single} solve of an unshifted system, i.e.\ ($\bs{A} + 0 \cdot\bs{I})^{-1} \bs{w}$ \citep{freund1993solution, clark2006rational, aune2013iterative, jegerlehner1996krylov}.
Alternatively, depending on the conditioning of $\bs{A}$, it might be preferable to solve all $N$ system separately, each with a different preconditioning approach \citep{simpson2008krylov}.

\section{Experiments}
In this section we apply our methodology to selected linear Gaussian models\footnote{Code used for the results in this section is available at \url{https://github.com/lellam/det_free_method}.}.
We begin with a simple toy model with a random precision matrix.
Our proposed methodology runs faster than a Cholesky based approach for large models, and scales to model sizes where Cholesky factors are not practical at all due to their memory costs.

\subsection{Random pattern precision matrices}
\begin{figure}[!ht]
\centering
\includegraphics[scale=.9]{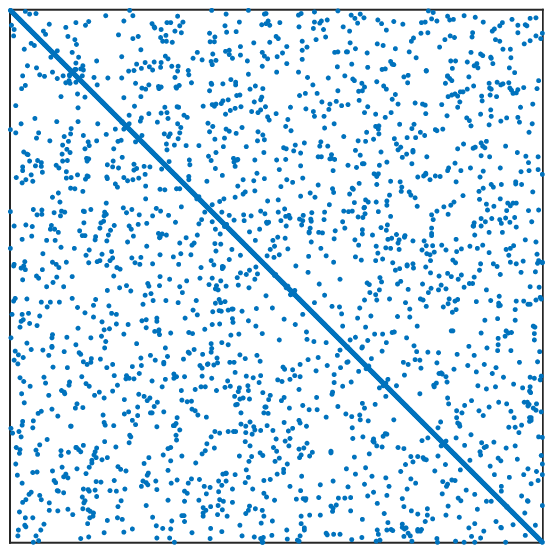}
\qquad
\begin{tabular}[b]{c c c c c}
	\toprule
	Matrix size &$10^3$&$10^4$&$10^5$ & $10^6$ \\
	\midrule
	\textbf{Standard\ MCMC} &  &  & \\ [0.5ex] 
	ESS& 1768 & 1529 &  NA & NA \\
	ESS/time & 6.9009 & 0.0362  & NA & NA \\
	
	\midrule
	\textbf{Our method} &  &  & \\ [0.5ex] 
	ESS& 883 & 543&  613 & 1271\\
	ESS/time & 7.6241 &  0.9460  &   0.3159 & 0.0558 \\
	\bottomrule
	\medskip
\end{tabular}
\captionlistentry[table]{A table beside a figure}
\captionsetup{labelformat=andtable}
\caption{\textbf{Left:} Sparsity pattern of a well conditioned randomly generated $10^3 \times 10^3$ matrix. \textbf{Right:} Comparison of MCMC efficiency for $\ln \gamma\!=\!-3$ for a number of different sized matrices.}
\label{fig:rand_mat}
\end{figure}

Consider the following model
\begin{equation*}
\bs{y} \sim \mathcal{N}(\bs{0},\, (\gamma^{-1} \bs{Q} + \gamma \bs{I})^{-1}),
\end{equation*}
where $\bs{Q}$ is a random, symmetric and positive definite matrix, generated by first generating a random Jacobi rotation of a positive definite diagonal matrix with elements in $[-\nicefrac{1}{2}, \nicefrac{1}{2}]$. 
By adding elements to the diagonal, the precision $\gamma^{-1}\bs{Q}+\gamma \bs{I}$ is diagonally dominant and therefore its condition number is bounded.
Each random precision matrix has $\sim\!3n$ non-zero elements;
an example is illustrated in Figure~\ref{fig:rand_mat}.
We use $\ln \gamma=-3$ to generate the data for this example. 

To perform inference,  we use a log-uniform prior on $\gamma$.  We run chains of length 10,000 for each example and tune the random walk proposal to obtain an acceptance rate of $20$--$40\%$.  We initialize the chain near its true value for this example.  We use $20$ terms in the rational approximation and run the shifted family conjugate gradient solver \citep{simpson2008krylov, jegerlehner1996krylov} to solve all linear systems in \eqref{eq:rational_approximation} in a single run.
The covariance matrices in this example are well-conditioned, so the conjugate gradient method converges rapidly without the help of a preconditioner.

We compare our proposed determinant-free method to a standard random walk MH sampler where we evaluate the model likelihood using a Cholesky factorisation.
As the table in Figure~\ref{fig:rand_mat} shows, the Cholesky based approach produces more independent samples for a fixed number of MCMC iterations, as measured by effective sample size (ESS)\@.
It is expected that dependencies between the model parameters and auxiliary variables will cause the Markov chain to mix slower.
Computing Cholesky factorisations requires more time per iteration and scales worse with problem size, which means that our scheme produces more effective samples per unit time.
For $n\ge 10^5$ a standard laptop doesn't have enough memory to run the Cholesky based scheme at all, while our method scales to much larger problems.

\subsection{Sparse covariances for spatial modelling of anti-social crime data}
\begin{figure}[!ht]
\centering
\includegraphics{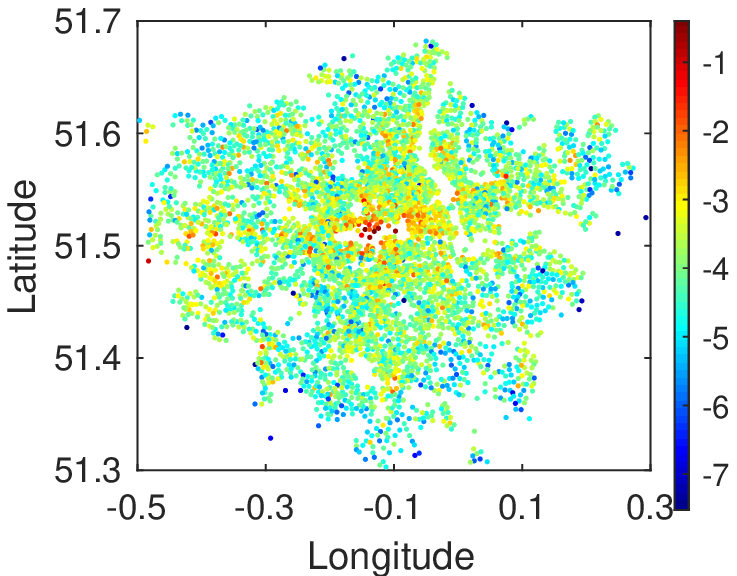}
\qquad
\begin{tabular}[b]{c c c c c}
	\toprule
	Parameter & $s^2$ & $l$ & $\tau$  \\
	\midrule
	\textbf{Std.\ MCMC} &&&&\\
	Mean & 0.13  & 0.06& 2.11 \\
	CD $\times 10^{-2} $  & 1.60 & 0.57 &4.91 \\
	ESS& 868 & 930& 956 \\
	ESS/time $s^{-1} \times 10^{-2}$ & 5.48 & 5.87 & 6.03 \\
	Time $s \times 10^4 $ & 1.59 & 1.59 & 1.59  \\
	\midrule
	
	\textbf{Det-free method} &&&&\\
	Mean & 0.16 & 0.06 & 2.10 \\ 
	SD  $\times 10^{-2} $ & 2.33 & 0.54 & 4.61 \\ 
	ESS& 213 & 242 & 340 \\ 
	ESS/time $s^{-1} \times 10^{-2}$  & 1.72 & 1.96 & 2.75  \\ 
	Time $\times 10^4 $& 1.24 & 1.24 & 1.24 \\ 
	\bottomrule
\end{tabular}
\medskip	
\caption{\textbf{Top:} Crime dataset showing log-crime-rate for reported anti-social crimes.  \textbf{Left:} Comparison of MCMC efficiency of standard MCMC and the proposed method.}
\label{fig:crime}
\end{figure}

We begin by applying our method to model the spatial distribution of anti-social criminal activity in London, using count data obtained from the UK government website.  Log-crime-rates were calculated for each Lower Layer Super Output Area (LSOA), defined as number of crimes divided by LSOA population, and each crime-rate was assigned to the central location of the LSOA\@.  After pre-processing, our dataset consists of 4,826 log-crime-rates, one for each region.  LSOA data was obtained from the Office of National Statistics. Figure~\ref{fig:crime} (\textbf{top}) illustrates the dataset.

The classical geostatistical model \citep{gelfand2010handbook}, decomposes observations (here one-dimensional) of a spatial stochastic process at locations $\bs{s}\subseteq\mathbb{R}^2$ as  $y(\bs{s}) = \mu(\bs{s}) + \eta(\bs{s}) + \epsilon(\bs{s}),$  where $\mu(\bs{s})$ is a deterministic mean function, $\eta(\bs{s})$ is a continuous zero-mean stochastic process and $\epsilon(\bs{s})$ is white noise.

Our choice of mean function $\mu(\bs{s})$ is $\mu(\bs{s}) = \beta_0 + \sum_{h=1}^5 \beta_h \exp \big\{ -\frac{1}{2\sigma_h^2} {\big\| \bs{s} - \bs{s}_h \big\|_2^2 } \big\}$,  where the constant $\beta_0$ represents the background crime-rate and the radial basis functions capture the general trend of an increase in crime in built-up areas.
The five coefficients $\{ \beta_h \}_{h=1}^5$ are held fixed at their maximum-likelihood values, after fitting a linear regression model.
The radial basis functions are centred at $\bs{s}_h$ found using the k-means algorithm and the scaling $\sigma_h$ is set equal to the smallest pairwise distance between the $\bs{s}_h$.
Centering the GP around the linear combination of basis functions focusses the model's attention on deviations from the general trend.  The stochastic process $\eta(\bs{s})$ is expected to capture localised crime `hot-spots'.
Those tend to only influence surrounding neighbourhoods -- a phenomenon that can be appropriately modelled with a compactly supported Wendland kernel \citep{Rasmussen06gaussianprocesses}:
\begin{equation*}
\textstyle
k_{\bs{\theta}}(\bs{x}_i, \bs{x}_j)  = s^2 \Big(1 -\frac{\|  \bs{x}_i -  \bs{x}_j \|_2}{l} \Big)^4_+ \Big( \frac{4\|  \bs{x}_i -  \bs{x}_j \|_2}{l} + 1 \Big).
\end{equation*}

We use independent weakly informative log-normal priors on $p(\tau)$, $p(s)$ and $p(l)$. The posterior over all unknown parameters $\bs{\theta}=(\tau, s, l)$ is then
\begin{equation*}
p(\tau, s, l | \bs{y}) \propto p(\bs{y}| \bs{\theta}) p(\tau)p(s)p(l).
\end{equation*}
We explore the posterior using a standard random walk MH both with determinant computations and our proposed augmented scheme.
We run chains of length 10,000 with a burn-in of 3,000.
The results and predictions are summarized in Figure~\ref{fig:crime}.
Due to the relatively small size of the dataset, the standard MCMC approach here produces slightly more independent samples per unit time -- while our method runs significantly faster, it mixes slower due to the augmented sampling space.

We expect that
our determinant-free method will provide more effective samples per unit time on
larger systems, with a greater spatial extent.
To test that idea we generated $n\!=\!10^4$ data locations from the uniform distribution over $[-0.5, 0.3]\times[51.25, 51.75]$ and sampled from the predictive distribution with the covariance parameters set to their posterior means.  We ran the above procedure for a chain of length $1,000$ and report the results for the model parameter $s^2$.
The ESS/time measured in $s^{-1} \times 10^{-4}$, for standard MCMC and our method was $9.2$ and $78.6$ respectively. Our method produced an order of magnitude more effective samples per unit time.

\subsection{Gaussian Markov random field models specified by a whitening matrix }
We now present a case where the model's precision matrix is known and is specified by a whitening matrix $\bs{L_\theta}$
\begin{align}
\label{whitening}
& \bs{L_\theta}\bs{x} = \bs{w}, \quad \bs{w} \sim N(\bs{0}, \bs{I}), \\
& \bs{Q}_{\bs{\theta}} = \bs{L_\theta}^T \bs{L_\theta},
\end{align}
This general setting is applicable for several models of interest, such as stochastic partial differential equation models discussed in ~\cite{kaipio2006statistical, lindgren2011explicit}.  In this case, the GMRF has sparse precision $\bs{Q}_{\bs{\theta}} = \bs{\Sigma_{\theta}}^{-1}$, and the latent process in \eqref{eq:latent_variables} is $\bs{x}\sim\mathcal{N}(\bs{\mu_\theta}, \bs{Q_\theta}^{-1})$.  We may simulate from the centered latent process by drawing white noise and solving the sparse linear system in~\eqref{whitening}; the Krylov methods that raise concerns over conditioning are not required in this case.  The covariance of the marginal likelihood in \eqref{eq:marginal_likelihood} is $\bs{S_\theta}=\tau^{-1}\bs{I} + \bs{A}\bs{Q}^{-1}_\theta\bs{A}^\top$, which is not necessarily sparse.  To evaluate the marginal likelihood without having to work with non-sparse matrices we can use the matrix inversion lemma:
\begin{equation}
\label{GMRF_like2}
\begin{aligned}
\ln p(\bs{y} | \theta) = & \frac{1}{2} \Big\{ \ln | \bs{Q}_\theta | + n\ln \tau - \ln |\bs{Q}_\theta + \tau \bs{A}^\top \bs{A}|  - \tau \bs{y}^\top \bs{y} \\& + \tau^2 \bs{y}^\top \bs{A} (\bs{Q}_\theta + \tau \bs{A}^\top \bs{A})^{-1} \bs{A}^\top \bs{y}  \Big\}  + \text{const}.
\end{aligned}
\end{equation}
For moderate scale models, the log-determinants in \eqref{GMRF_like2} can be evaluated with 2 Cholesky decompositions.

\begin{figure}[!ht]
\centering
\includegraphics[scale=.74, clip,trim=0cm -1cm 0cm 0cm]{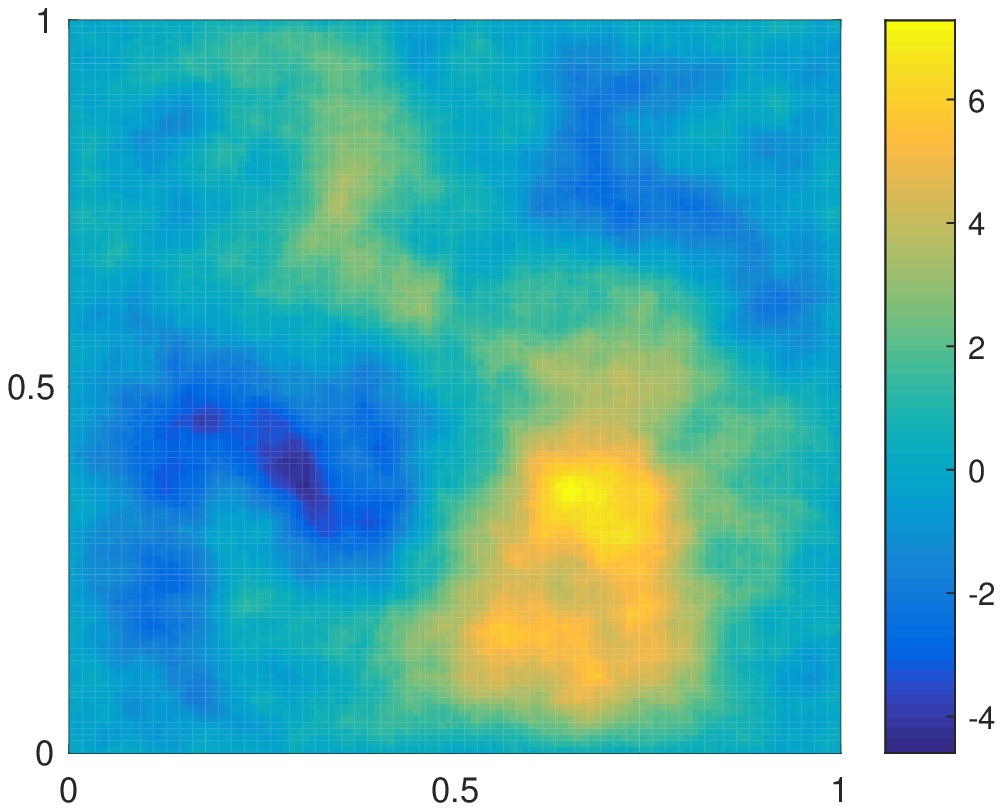}
\qquad
\begin{tabular}[b]{c  c c  c c}
	\toprule
	Grid  & \multicolumn{2}{c}{$120 \times 120$}& \multicolumn{2}{c}{$200 \times 200$} \\
	\toprule
	Case & $\ln \tau$ & $\ln \gamma$&  $\ln \tau$ & $\ln \gamma$ \\
	\midrule
	{\textbf{Std MCMC}}&&&&\\
	Mean&-1.99& -1.05 & -1.98 & -0.99 \\
	SD &0.008& 0.050& 0.011 & 0.037 \\
	ESS&1078& 1373& 796 & 1567 \\
	ESS/time & 0.387 & 0.493 & 0.053 & 0.105 \\
	\midrule
	{\textbf{Det-free MCMC}}&&&&\\
	Mean&-1.99 & -1.05 & -1.98 & -0.99 \\
	SD&   0.012& 0.068& 0.015 &  0.053 \\
	ESS&512& 636& 414 & 786 \\
	ESS/time & 0.238 &  0.295&0.067&0.128  \\
	\bottomrule
	\medskip
\end{tabular}
\vspace{-.5cm}
\captionlistentry[table]{A table beside a figure}
\captionsetup{labelformat=andtable}
\caption{\textbf{Left:} The true latent process for the $120 \times 120$ grid.   \textbf{Right:} Comparison of the efficiency of standard MCMC and the proposed method for $120 \times 120$ grid (with $15,000$ observations) and $200 \times 200$ grid (with $10,000$ observations).  Observations were drawn at random locations by interpolating the latent process and adding scaled white noise. }
\label{fig:gmrf_data_results}
\end{figure}

In this example, specify the precision by a smooth latent process: $\bs{L_\theta} = \frac{1}{\gamma} \bs{L}_D$, where $\bs{L}_D$ is a discrete approximation of the Laplacian with Dirichlet boundary conditions generated from the mask
\begin{equation*}
\begin{bmatrix}
& -1 & \\
-1 & 4 & -1 \\
& -1 & \\
\end{bmatrix}.
\end{equation*}   We define the latent process over a uniform quadrilateral mesh and consider the cases with $120 \times 120$ nodes and $200 \times 200$ nodes, so that $m = 14,400$ and $m=40,000$ for each case, respectively. We generate synthetic datasets of size $n$ at locations drawn uniformly over $[0, 1]^2$.  The matrix $\bs{A}$ performs linear interpolation to estimate the latent process at the observation locations.  We use the values $\ln \tau = -2$ and $\ln \gamma = -1$ to generate the true latent process in both cases and the latent process for the $120 \times 120$  case is presented in Figure~\ref{fig:gmrf_data_results}.   We perform inference on the marginal posterior for $\gamma$ and $\tau$ using log-uniform priors for each.

Our preliminary runs revealed that $\bs{Q}_\theta$ is poorly conditioned, e.g.\ the $120 \times 120$ example had a condition number of $5.57 \times 10^7$, which is expected to increase with $m$.  Unlike for models specified by covariance, Krylov methods must be performed directly on the latent process and the observation noise does not help with conditioning as it does for standard GP models.  As a result, Krylov methods are expected to require a large number of iterations.  In this setting, we make use of the sparse linear system in~\eqref{whitening}, which can be solved using a banded linear system solver.

We again explore the posterior using a standard random walk MH: first with determinant computations, using $2$ Cholesky decompositions per iteration; and then with our proposed augmented scheme.  We run chains of length $10,000$ and initialize the parameters close to their true values.  The results in Figure~\ref{fig:gmrf_data_results} show that, for the both examples, our method out-performs the standard MCMC approach in cost per iteration.  For large scale problems, our method can outperform standard MCMC in ESS/time, despite the slower mixing caused by adding a larger number of auxiliary variables to the model.

\section{Discussion}
We have shown how to introduce auxiliary random variables to replace the determinant computation arising in Gaussian models whose covariance is specified by unknown model parameters. The Markov chains for our method mix more slowly per iteration than a Cholesky-based approach, due to the additional auxiliary variables, but can be much cheaper per iteration.  Our method can be fast because it exploits fast matrix-vector operations, such as for models specified by sparse matrices, and it never creates dense matrices. These properties are particularly beneficial when the Cholesky decomposition is prohibitively expensive due to the fill-in, or where the Cholesky decomposition is expensive to obtain.

In practice, latent functions require a degree of smoothness, which forces the smallest eigenvalue in a large system close to zero and can result in a poorly conditioned system.
The irony is that a system being poorly conditioned implies that it is almost low rank.
Low rank systems are more constrained than an arbitrary process, and so should be less expensive to work with.
In our crime example we exploited the low rank structure and worked with a reasonably well conditioned covariance matrix.
However, Krylov methods do not work as well with extremely poorly-conditioned precision matrices as found in some GMRFs.
In future work we will explore ways to exploit low rank structure more generally, so that our auxiliary variable scheme can focus on exploring the degrees of freedom of the process that have significant posterior uncertainty.
We are also exploring applications of the methodology to inverse problems arising in partial differential equation models\looseness=-1.

\section*{Acknowledgements} L.E. was supported by the EPSRC.  H.S. was supported by the Gatsby Charitable Foundation.   M.G. was supported by EPSRC [EP/J016934/1, EP/K034154/1, EP/P020720/1] and an EPSRC Established Career Fellowship.  This work was conducted when L.E. and I.M. were visiting the Alan Turing Institute, London.

\bibliographystyle{wb_stat}
\bibliography{ref}

\end{document}